\documentclass[twocolumn]{aastex62}
\usepackage{hyperref}
\usepackage{bm}

\newcommand{\vth}{v_{\text{th}}}
\newcommand{\taul}{\tau_L}
\newcommand{\kapl}{\kappa_L}
\newcommand{\XSTAR}{\textsc{\scriptsize XSTAR}}
\newcommand{\CLOUDY}{\textsc{\scriptsize CLOUDY }}
\newcommand{\Athena}{\textsc{Athena\scriptsize ++ }}
\newcommand{\angstrom}{\mbox{\normalfont\AA}}
\received{December 2018}
\revised{July 2019}
\accepted{July 2019}
\submitjournal{ApJ}
\shorttitle{Photoionization calculations of the radiation force due to spectral lines in AGNs}
\shortauthors{Dannen et al.}
\begin{document}
\title{Photoionization calculations of the radiation force due to spectral lines in AGNs}
\correspondingauthor{Randall Dannen}
\email{randall.dannen@unlv.edu}
\author[0000-0002-5160-8716]{Randall C. Dannen}
\affiliation{Department of Physics \& Astronomy \\
University of Nevada, Las Vegas \\
4505 S. Maryland Pkwy \\
Las Vegas, NV, 89154-4002, USA}
\author[0000-0002-6336-5125]{Daniel Proga}
\affiliation{Department of Physics \& Astronomy \\
University of Nevada, Las Vegas \\
4505 S. Maryland Pkwy \\
Las Vegas, NV, 89154-4002, USA}
\author[0000-0002-5779-6906]{Timothy R. Kallman}
\affiliation{NASA Goddard Space Flight Center \\
Greenbelt, MD 20771, USA}

\author[0000-0002-5205-9472]{Tim Waters}
\affiliation{Theoretical Division, Los Alamos National Laboratory}
\begin{abstract}
One of the main mechanisms that could drive mass outflows in AGNs is radiation pressure due to spectral lines. Although straightforward to understand, the actual magnitude of the radiation force is challenging to compute because the force depends on the physical conditions in the gas, and the strength, spectral energy distribution (SED), and geometry of the radiation field.  We present results from our photoionization and radiation transfer calculations of the force multiplier, $M(\xi,t)$, using the same radiation field to compute the gas photoionization and thermal balance. We assume low gas density ($n = 10^4~\rm{cm^{-3}}$) and column density ($N \leq 10^{17}~\rm{ cm^{-2}}$), a Boltzmann distribution for the level populations, and the Sobolev approximation.  Here, we describe results for two SEDs corresponding to an unobscured and obscured AGN in NGC 5548.  Our main results are the following:  1) although $M(\xi,t)$ starts to decrease with $\xi$ for $\xi \gtrsim 1$ as shown by others, this decrease in our calculations is relatively gradual and could be  non-monotonic as $M(\xi,t)$ can increase by a factor of few for $\xi \approx 10-1000$; 2) at these same $\xi$ for which the multiplier is higher than in previous calculations, the gas is thermally unstable by the isobaric criterion; 3) non-LTE effects reduce $M(t,\xi)$ by over two orders of magnitude for $\xi \gtrsim 100$.  The dynamical consequence of result (1) is that line driving can be important for $\xi$ as high as $1000$ when the LTE approximation holds, while result (2) provides a natural cloud formation mechanism that may account for the existence of narrow line regions. Result (3) suggests that line driving may not be important for $\xi\gtrsim100$ in tenuous plasma.
\end{abstract}
\keywords{
galaxies: active - 
methods: numerical - 
hydrodynamics - radiation: dynamics
}
\section{Introduction} \label{sec:intro}
Active galactic nuclei (AGNs) are examples of astrophysical objects that are extremely luminous radiation sources over a broad range of energies. A consequence of such efficient conversion of gravitational binding energy is the production of outflows in the form of relativistic jets and winds of ionized gas with velocities approaching a few percent the speed of light \citep[e.g.,][]{Kaspi02,Cetal02, Tetal10, Reeves18}.  Radiative and mechanical AGN feedback may play an important role in the evolution of the host galaxy \citep{Silk98,FurLoeb03, Haiman06, Hopkinsetal06a, Ciotti10, McCarthy10, Ostriker10,Fabian12, FG12, Choietal14}, and the winds in particular are responsible for imprinting a host of spectral features in the optical, UV, and X-ray bands.  These features take the form of broad absorption lines, warm absorbers and ultra-fast outflows \citep{Wetal91, RF95, Cetal03, Getal11, Hetal13, Ketal14, Netal15, Metal17,Aetal18}, and much work is still required to determine the mechanism by which these outflows form.  
\par
Plasma codes such as \XSTAR\footnote{\url{https://heasarc.nasa.gov/lheasoft/xstar/xstar.html}}~\citep{KB01} and \CLOUDY\footnote{\url{https://www.nublado.org/}} \citep{FPvH13} can be used to model these features, as they perform detailed photoionization, radiative transfer, and energy balance calculations.  {Several groups have begun incorporating these calculations into hydrodynamical codes, and this is currently the most accurate approach for comparing theory with observations \citep[e.g.,][]{Salz15,RVetal16,Kinch16,Detal17, Higginbottom17}.}\par
{Still, even within this modeling framework, a full treatment of the effects of radiation on gas dynamics requires many time consuming calculations, so various further approximations are made.  For instance, most theoretical studies of gas flows that include thermal driving neglect or simplify the treatment of the radiation force, $\bm{f}_{\rm rad}$, whereas studies of radiation pressure driven flows typically neglect or simplify the radiative heating and cooling rates.
In some applications, one could justify ignoring the radiation force by referring to the rule of thumb that ``radiation can heat (cool), but frequently finds it difficult to push'' \citep{Shu92}.  However, under some circumstances, radiation can effectively push gas, such as when the total opacity of the gas, $\kappa_{\rm tot}$, is dominated by the contribution from photon scattering (thereby providing no energy transfer) rather than from photon absorption (which mainly provides heating).  OB stars and cataclysmic variables are examples of objects where the total opacity in their upper atmospheres and winds is dominated by contributions from spectral line transitions, which mostly scatter photons,  hence their winds are driven by the so-called line force, $f_{\rm rad, l}$ \cite[see][CAK hereafter]{CAK}.  Early theoretical work suggested that the radiation force can also be the main mechanism driving supersonic gas flows in AGNs \citep[]{Mushotzky72,AL94, Metal95, PSK00, P07}, and numerous observations appear to confirm these expectations \citep{Foltz87, Srianand02, Ganguly03, Gupta03, North06, Bowler14, Lu18,MasRibas19}}.\par
{This paper is focused on developing the capability to self-consistently model situations where both the momentum and energy of the radiation field are dynamically important.  Consider for instance flow regimes where the rate of work done by the radiation force,  $\bm{f}_{\rm rad} \cdot \bm{v}$, where $\bm{v}$ is the flow velocity, transitions from being smaller than the rate of heat deposition ($Q \equiv -\mathcal{L} = {\Gamma - \Lambda}$, where $\Gamma$ and $\Lambda$ denote the net radiative heating and cooling rates, respectively) to larger, or vice-versa.  In such regimes, it is not accurate to perform hydrodynamical calculations without the detailed source terms that can be calculated from photoionization codes.  Likewise, photoionization calculations alone do not suffice to model the spectra, as the gas dynamics is playing a crucial role in determining the spectral features.  This situation is expected, for example, in regions of disk winds where heating is responsible for the initial mass loading in the subsonic part of the flow, but becomes less important in the supersonic regions once the gas temperature does not change much \citep[e.g.,][D17 hereafter]{Detal17}.  The gas can then remain nearly isothermal while still undergoing acceleration due to the line force (see also Dannen \& Proga, in preparation).}  \par
\begin{figure*}
  \centering
  \includegraphics[width=0.92\textwidth]{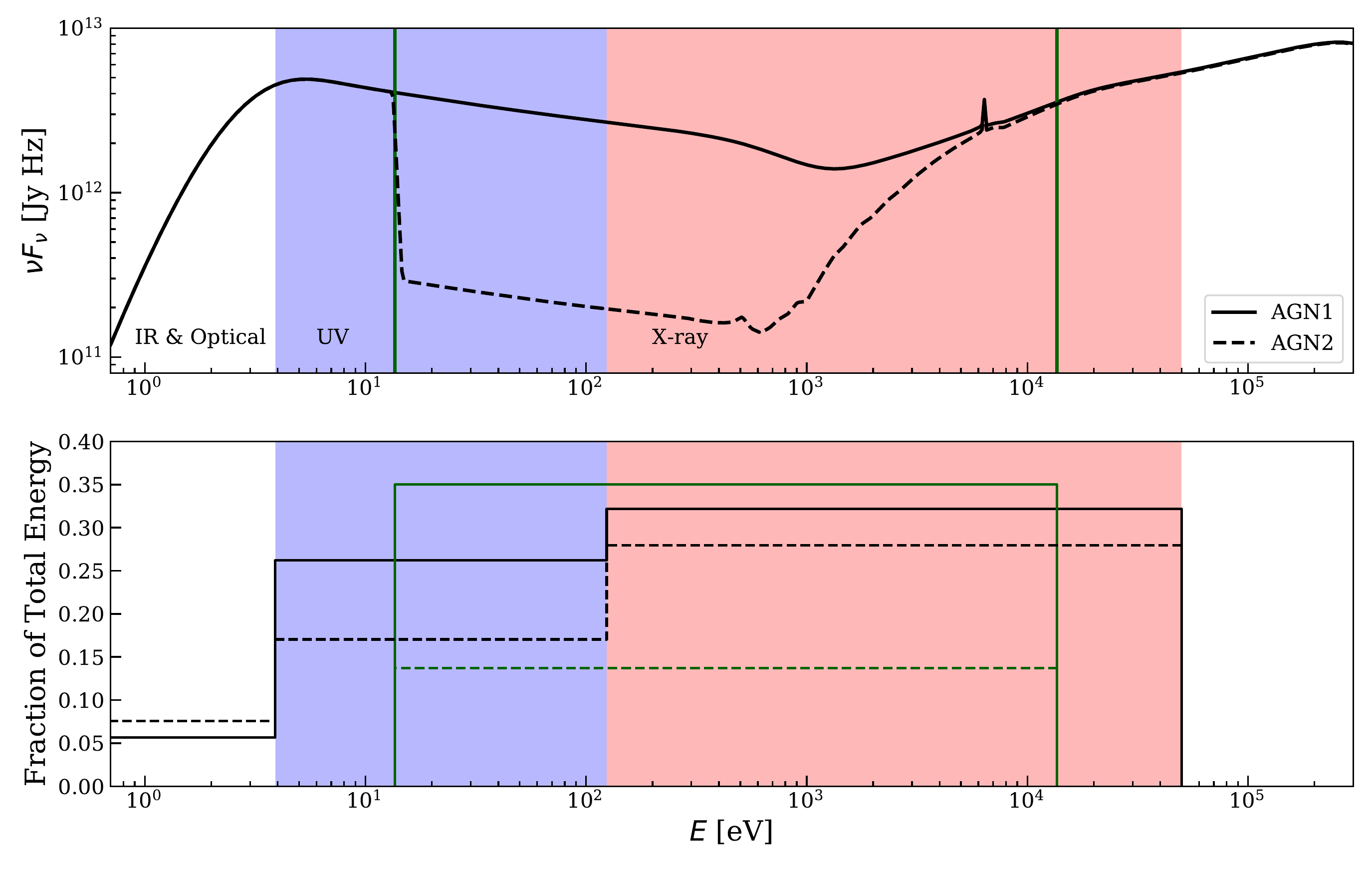}
  \caption{{\it Top panel:} Two representative AGN SEDs used in our calculations: AGN1 represents the unobscured SED (solid line) while AGN2 represents the obscured SED  (dashed line) of the AGN in NGC 5548.  We use different color shading to show the UV and X-ray energy bands. The two vertical green lines mark the energy interval used to calculate $\xi$ (13.6~eV -- 13.6~keV).  {\it Bottom panel:} Fraction of the total energy of the SEDs in each energy component band for AGN1 (solid lines) and AGN2 (dashed lines). We also show the fraction of the total energy used in computing~$\xi$ (i.e., the fraction of the total energy between the vertical green lines in the top panel).}
  \label{fig:seds}
\end{figure*}
D17 described a general method for the self-consistent modeling of the outflow hydrodynamics that results from the irradiation of optically thin gas by a radiation field with an arbitrary strength and SED. D17 used  the photoionization code \XSTAR~to calculate  ${\Gamma}$ and ${\Lambda}$ as a function of  gas ionization parameter
\begin{equation}
\xi = \frac{4\pi F_{X}}{n_H},
\end{equation}
and gas temperature, $T$, where $F_X$ is the integrated flux from 0.1~Ry--1000~Ry and $n_H$ the hydrogen nucleon number density.  D17 explored several SEDs: those of Type 1 and Type 2 AGN from Fig.~\ref{fig:seds}, as well as SEDs for hard and soft state X-ray binaries, bremsstrahlung and blackbody. This general method was applied to study the hydrodynamics of 1-D spherical winds heated by a uniform radiation field using the magnetohydrodynamic (MHD) code \Athena \citep{GS05, GS08, Athena08}. \par
D17 found that in all the stated cases a wind settles into a transonic, steady state.  The wind is at near radiative heating equilibrium (i.e., $\mathcal{L} \simeq 0$) until adiabatic cooling becomes important and the flow  temperature can be significantly smaller  than the one corresponding to the radiative equilibrium value for a given $\xi$.  D17's  results also show how the efficiency with which the radiation field transfers energy to the wind is dependent on the SED of the external source, particularly the relative flux of soft X-rays. This soft X-ray dependence is related to the thermal stability of the gas \citep{Field65}, namely a relative deficit of the soft X-rays leads to more unstable gas \citep[see also][]{KMc82} which in turn increases heat deposition and the flow velocity. Overall, D17's results demonstrate how detailed photoionization calculations are essential to properly capture the flow dynamics. \par
In this paper, we make the next step in our development of a self-consistent comprehensive model of astrophysical winds.  As in D17, we employ the photoionization code \XSTAR, this time to compute not only the heat deposition but also the line force as a function of $\xi$ and $T$.  Here we limit our analysis to the two AGN SEDs shown in Fig.~\ref{fig:seds}.  In upcoming papers, we plan to explore the parameter space of these calculations to identify regions where the radiation force dominates over thermal driving in determining the mass loss rate or terminal velocity (or both).
\par
The methods and results from this paper make it possible to self-consistently combine radiation driving and thermal driving due to the same radiation field. In \S{\ref{sec:num-meth}}, we detail our methodology for arriving at force multipliers given an arbitrary SED.  We present the results from our calculations in \S{\ref{sec:results}}, a discussion of these results and non-LTE effects in \S{\ref{sec:discussion}}, and a summary of our conclusions in \S{\ref{sec:conclusion}}.  The data tables for the AGN1 and AGN2 heating and cooling rates from D17 and the force multipliers presented in this work, along with sample code for utilizing them, can be found on our project webpage\footnote{\url{http://www.physics.unlv.edu/astro/xstartables.html}}. 
\par
\begin{figure*}
  \centering
  \includegraphics[width=\textwidth]{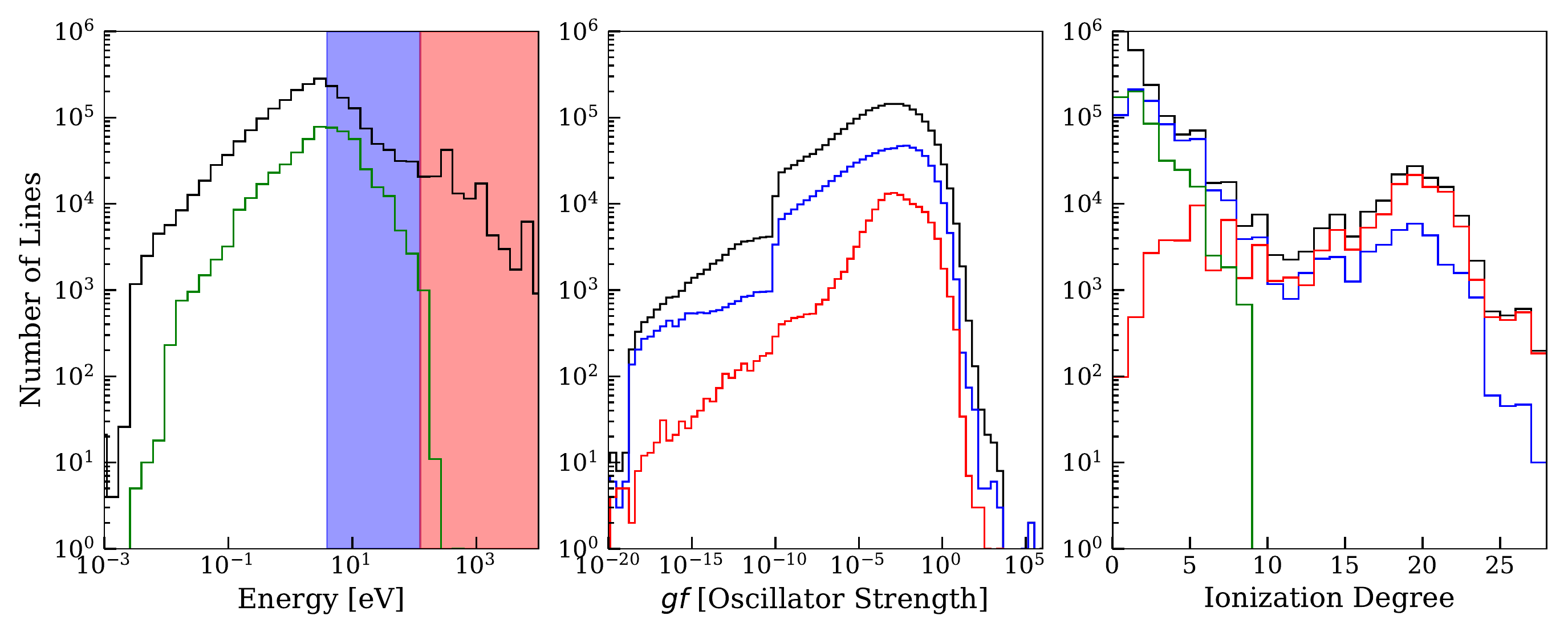}
  \caption{Histograms illustrating properties of the lines used in our calculations.  {\it Left panel:} Solid lines indicate the number of lines as function of energy in units of eV for the atomic database used in this work (black line) and used in SK90 (green line).  For emphasis, we shade in blue the energy range for UV photons and in red for the X-rays.  {\it Middle panel:} Number of lines as a function of oscillator strength $gf$ for the entire line list (black line), including only the UV lines (blue line), and only X-ray lines (red line).  {\it Right panel:} Number of lines as a function of ionization degree for the atomic database used in this work (black line), UV lines only (blue line), X-ray lines only (red line), and the atomic database used in SK90 (green line).}
  \label{fig:line-list}
\end{figure*}
\section{Methodology} \label{sec:num-meth}
In this section, we describe our process for calculating the force multiplier.   We base our method on the CAK formalism. More specifically, we follow \cite[SK90 heareafter]{SK90} who used \XSTAR~ version 1 to calculate the state of the gas.  Here we use \XSTAR~ version 2.37 along with the most up-to-date lists of spectral lines, using the AGN SEDs from Fig.~\ref{fig:seds} as inputs to the code. In addition, we use the same radiation field to compute both $f_{\rm rad, l}$ and $\mathcal{L}$.  \par
\subsection{Underlying SEDs}
This paper builds upon several previous studies of AGN gas flows where radiation source terms are included in a progressively more self-consistent manner \citep[e.g.,][]{PSK00, P07, PW15, Detal17}.  To properly include both radiation heating/cooling and radiation forces, one needs to accurately compute the coupling between radiation and matter, i.e. the gas opacity and emissivity, from the underlying spectral energy distribution (SED) of the electromagnetic radiation.  The top panel of  Fig.~\ref{fig:seds} shows examples of unobscured and obscured AGN SED of NGC 5548 adopted from \cite{Metal15}.\par
In the bottom panel of Fig.~\ref{fig:seds}, we mark the fraction of the total energy contained in various portions of the electromagnetic spectrum. We note that the comparison of ratios between the hard and soft UV and X-ray fractional energy is likely an important factor.  That is, the quantity of the radiation is clearly important, but the shape of the radiation field is also important. For example, modeling of AGN observations requires spectra that are of a certain hardness \citep[e.g.,][and references therein]{Metal15}.  However, only a fraction of the physically relevant SED energy range can be observed.  The behavior in the unobservable EUV region, for instance, must be assumed.  Additionally, the SED also affects the gas thermal stability \citep[e.g.,][]{KMc82,K99, Metal15, Detal17}.  In particular, AGN1 and AGN2 both have regions of isobaric thermal instability; we return to this point in \S{\ref{sec:results}}.  The deficit of soft photons in AGN2 also allows for isochoric instability, which leads to the formation of non-isobaric clouds \citep{WP19}.  In this context, the AGN is being obscured or not obscured from point of view of the gas and not necessarily the observer.\par
{We present these two SEDs since there is evidence that the evolution of the SED and therefore the gas dynamics may be due to material moving between the AGN and our line of sight \citep[LOS; e.g.,][]{Capellupo11,Capellupo12}.  AGN1, the unobscured SED, is the intrinsic SED of the AGN and AGN2, the obscured SED, represents the AGN SED through a column density of material $N_H = 1.45\times10^{20}~{\rm cm}^{2}$ \citep{Metal15}.  This allows us to consider situations where the SED incident on the gas experiences the attenuated SED}.  \par
\subsection{Atomic Line Lists}
SK90 considered an X-ray binary system where black body radiation from a star drives a stellar wind which was irradiated by X-rays emitted by a companion. Unlike SK90, we assume the radiation field for both the ionizing flux and line driving is the same, as this is more appropriate for modeling gas dynamics in AGNs.  Other authors have also explored the line force due to AGNs \citep[e.g.][]{AL94, CN03,E05, Cetal09, SC11} but they used a different photoionization code (e.g. \textsc{\scriptsize CLOUDY}) and different atomic data sets and line lists. \par
The line list that we use is a combination of the \XSTAR~ atomic data set and the atomic data curated by Robert L. Kurucz\footnote{\url{http://kurucz.harvard.edu}}.  We take special care when merging these atomic data sets to not double count any lines.  If a line was found in both data sets, we prioritize the \XSTAR~ atomic data for X-ray lines and high energy UV lines and Kurucz's data set otherwise.  Information about the distribution of lines as a function of energy, oscillator strength, and ionization degree is shown in Fig.~\ref{fig:line-list}.  This figure also includes information about the atomic data set used by SK90.  Our current atomic data set contains over two million lines, covering a wider range of energies and ionization degrees and allowing for a more complete calculation of the force multiplier compared to previous studies, especially due to X-rays lines and lines from highly ionized plasma.
\subsection{Force Multiplier Calculations}\label{sec:fmult-calc}
\begin{figure*}
  \centering
  \includegraphics[width=\textwidth]{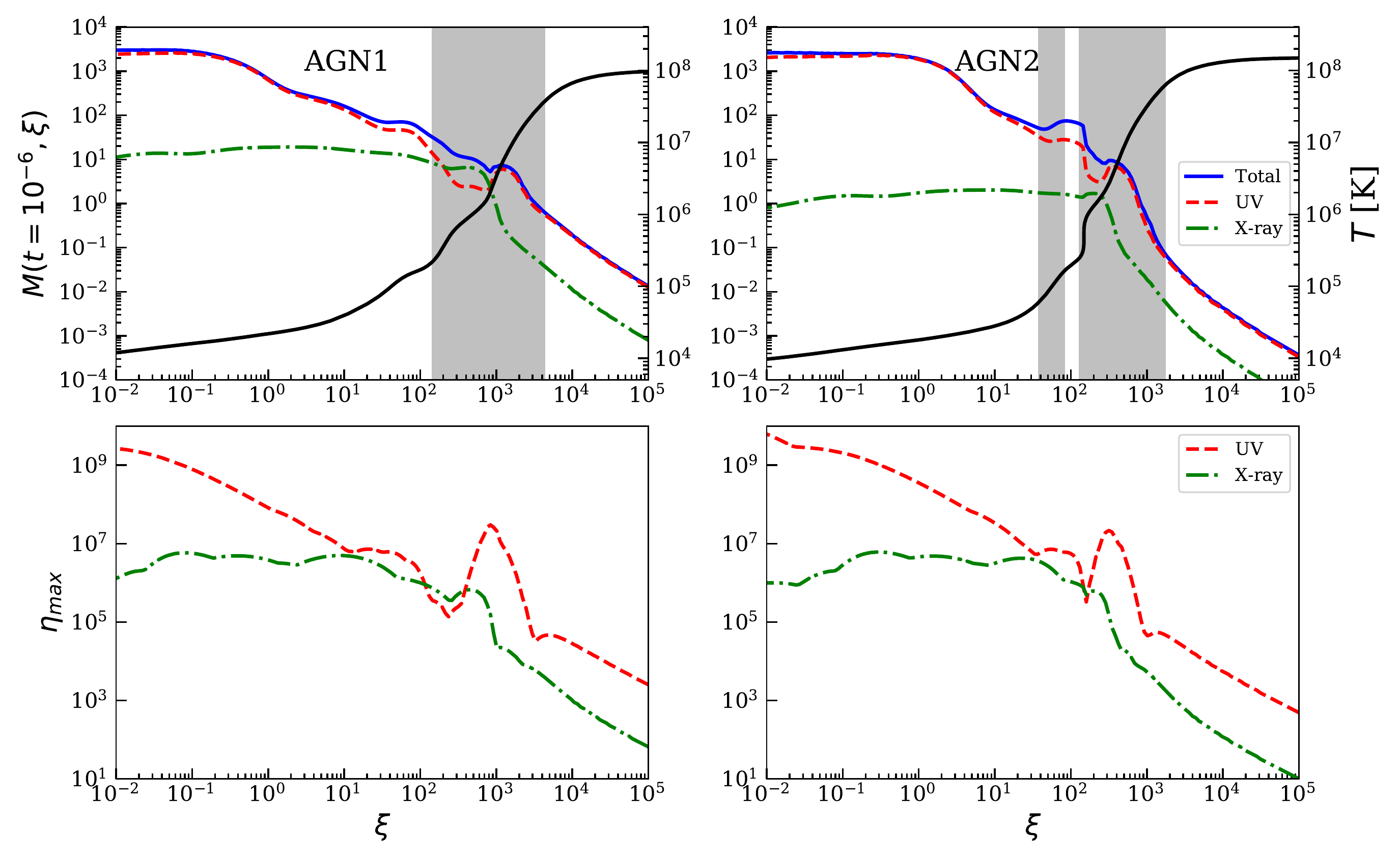}
  \caption{Summary of results from our photoionization and line force calculations.
  {\it Top panels:} The force multiplier $M(\xi,t)$ as a function of $\xi$ for an optically thin gas with small optical depth parameter, $t=10^{-6}$, for AGN1 (left) and AGN2 (right). This small value of $t$ yields a proxy for $M_{\rm max}$.  Plotted against the left vertical axes are values of $M_{\rm max}$ due to all lines (solid blue curve), UV lines only (dashed red curve) and X-ray lines only (dash-dotted green curve). Plotted against the right vertical axes is the {photoionzation equilibrium temperature} determined by \XSTAR~ corresponding to $\mathcal{L}=0$ (solid black line).  The shaded regions indicate the parameter space where the gas is thermally unstable by the isobaric criterion (i.e., $\left[\partial \log T / \partial \log \xi\right]_{\mathcal{L}} \geq 1$; \citep[e.g.,][]{Barai12}).  {\it Bottom panels:} The opacity of the single strongest UV and X-ray line as a function of $\xi$ (dashed red and dash dotted green curves, respectively).  The line opacity is in units of electron scattering opacity (see Eq.~\ref{eq:eta-def}).}
  \label{fig:mt_eta}
\end{figure*}

We adopt the same elemental abundances as \cite{MKK16} in both models.  The ionization balance is determined by the external radiation field rather than by the LTE assumption (i.e. Saha ionization balance), using the  photoionization code \XSTAR~ to determine the ion abundances as a function of $\xi$.  The gas temperature is also function of $\xi$, which entails an implicit assumption that the gas is optically thin. {To guarantee this this while also ensuring that collisional de-excitation processes are negligible compared to radiative processes in determining the ionization balance, we set the hydrogen nucleon number density in \XSTAR~ to~$n_H = 10^4~{\rm cm}^{-3}$.  This value is in accordance with density estimates for AGN outlfows \citep{Aetal18}.  The column density was set to $N=10^{17}~{\rm cm}^{-2}$, the luminosity to $10^{46}$~ergs~s$^{-1}$, and the ionization parameter at the inner radius to $\log (\xi) = 5$, placing the most ionized gas $\approx 1$~pc away from the source and least ionized gas $\approx 3000$~pc from the source.  This results in the range $\xi$ spanning $\log(\xi)$ from -2 to 5, where \XSTAR~ defines $\approx 160$ spatial zones with step sizes $\Delta \log(\xi) \approx 0.4$.}  In the two top panels of Fig.~\ref{fig:mt_eta}, we show the photoionization equilibrium temperature (i.e., the temperature for which the total amount of energy absorbed from the incident radiation field should balance the total emitted energy in lines and continua) as a function of $\xi$ (black solid lines) and we shaded regions where the condition for isobaric thermal instability is satisfied \citep[]{Field65}.
\par
\begin{figure*}
  \centering
  \includegraphics[width=\textwidth]{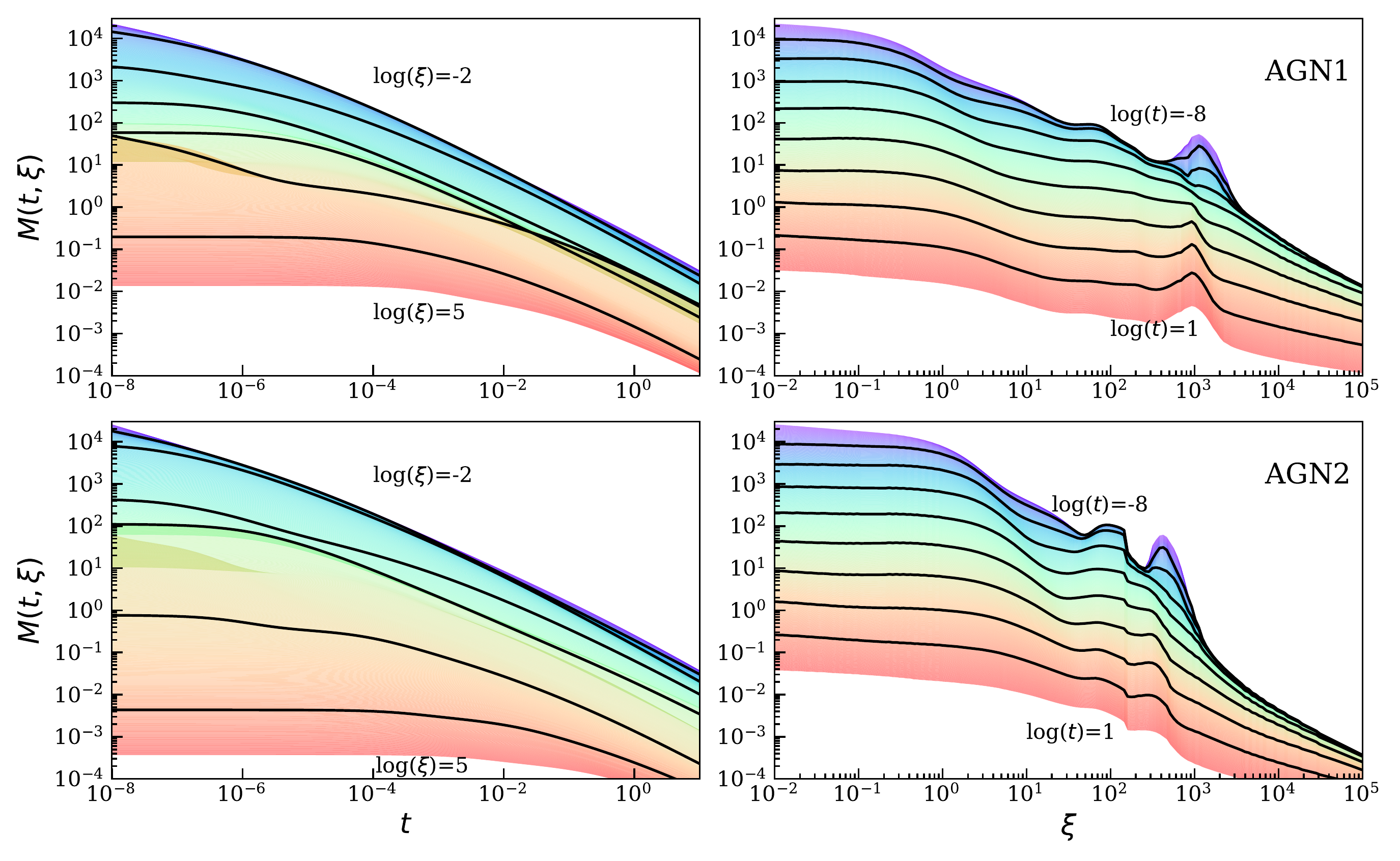}
  \caption{
  The main results from our photoionization and line force calculations. 
  {\it Top panels:}  These correspond to AGN1.  The left panel shows our results as a function of $t$ for a fixed $\xi$.  The solid lines show results for a fixed $\xi$ [from top to bottom, $\xi$ increases from $10^{-2}$ (violet region) to $10^5$ (red region)], and the $\xi$ levels are separated by $\Delta \log(\xi) = 1$.  The right panel shows results of our force multiplier calculations as function of $\xi$ for a fixed $t$.  The solid lines show results for a fixed $t$ [from top to bottom, $t$ increases from $10^{-8}$ (violet region) to $10$ (red region)], and the $t$ levels  are separated by $\Delta \log(t) = 1$.  {\it Bottom panels:}  Same as the top panels for AGN2.}
  \label{fig:summary-fig}
\end{figure*}
The force per unit mass due to an individual line can be computed as
\begin{equation} \label{eq:ind-line-force-1}
f_L = \frac{F_\nu \Delta\nu_D}{c}\frac{\kapl}{\taul}
\left(
1 - e^{-\taul}
\right),
\end{equation}
where $\kappa_L$ is the line's opacity, $\tau_L$ the line's optical depth, $\Delta \nu_D =\nu_0 \vth/c$ is the line's thermal Doppler width, with $\vth$ being defined as the thermal speed of the ion that a given atomic line belongs to, and $F_\nu$ is the specific flux \citep{C74}.  The optical depth for specific line in a static atmosphere is
\begin{equation} 
\tau_L = \int\limits_{r}^{\infty}\rho \kappa_L dr,
\end{equation}
while in an expanding atmosphere it is
\begin{equation} \label{eq:exp-optical-depth}
\tau_L = \rho \kappa_L l_{\rm Sob}.
\end{equation}
Here, $l_{\rm Sob}\equiv\vth/|dv/dl|$ is the so-called Sobolev length. 
CAK defined a local optical depth parameter
\begin{equation} \label{eq:t-def}
t = \sigma_e \rho\, l_{\rm Sob}.
\end{equation}
We take $\vth$ to be the proton thermal speed at 50,000 K.  For a given line, we can define the opacity accounting for stimulated emission,
\begin{equation}\label{eq:kappa_L}
\kappa_L = \frac{\pi e^2}{m_e c} gf \frac{N_L/g_L - N_U/g_U}{\rho \Delta \nu_D}, 
\end{equation}
where $\kappa_L$ is the opacity in units cm$^2$ g$^{-1}$ (where all other symbols have their conventional meaning).  Following SK90, we assume a Boltzmann distribution (i.e. the Local Thermodynamic Equilibrium, LTE, assumption) when determining the level populations.  For SK90's calculation, \XSTAR~ version 1 did not explicitly calculate the populations of excited levels, and also such a calculation was not computationally feasible at the time.  In our current work, the LTE assumption is made in order to allow the use of the Kurucz line list and to allow for direct comparison with CAK, SK90, \cite{Gayley95}, and \cite{Puls00}.  
This line list is more extensive in the optical and ultraviolet than the lines currently considered by \XSTAR.  However, the Kurucz line list does not include the associated collision rates or level information which are needed in order to calculate non-LTE level populations.  The LTE assumption results in a larger population of the excited levels when compared to a non-LTE calculation (see \S\ref{sec:discussion} for discussion on these points). \par
It is conventional to rewrite the optical depth as
\begin{equation} \label{eq:taul-rescale}
    \taul = \eta t,
\end{equation}
where $\eta$ is a rescaling of line opacity
\begin{equation} \label{eq:eta-def}
\eta = \kapl / \sigma_e.
\end{equation}
We can now write an expression for the total acceleration due to lines as
\begin{equation} \label{eq:line-acc-def}
a_L = \frac{\sigma_e F}{c} M(t,\xi),
\end{equation}
where $M(t,\xi)$ is the total force multiplier given by
\begin{equation} \label{eq:force-mult-def}
M(t,\xi) = \sum\limits_{\text{lines}} \frac{\Delta \nu_D \,F_\nu}{F} \frac{1}{t}
\left(
1 - e^{-\eta t}
\right).
\end{equation}
While $t$ depends on the temperature through $\vth$, this temperature value is arbitrary and has no direct role in $M(\xi,t)$; our calculations of $M(\xi,t)$ just require one to be specified  {(i.e. the $\vth$ terms cancel upon expanding the various components of Eq.~\ref{eq:force-mult-def}; see \citealt{Gayley95} for a detailed discussion of this point and an alternative formalism).}\par
CAK found that $M(\xi,t)$ increases with decreasing $t$ and it saturates as $t$ approaches zero (i.e., gas becomes optically thin even for the most opaque lines). We will refer to the saturated value of $M(\xi,t)$ as $M_{\rm max}$. CAK also showed that for OB stars, $M_{\rm max}$ can be as high as $\sim 2000$ \cite[see also ][]{Gayley95}.  This means that the gravity can be overcome by the radiation force even if the total luminosity, $L$, is much smaller than the Eddington luminosity, ${L_{\rm Edd}} = 4 \pi c G M / \sigma_{e}$.  In other words, the radiation force can drive a wind when $L'  M_{\rm max} >1$, where $L' \equiv L / L_{\rm Edd}$ is the so-called Eddington factor. \par
The key result of SK90 was to show how $M(\xi,t)$ changes not only as a function of $t$ but also as a function of $\xi$.  In particular, they found that $M_{\rm max}$ increases gradually from $\sim 2000$  to $\rm 5000$ as $\xi$ increases from 1 to $\thicksim 3$ and then drops to $\sim 0.1$ at $\xi \sim 1000$.  The line force becomes negligible for $\xi > 100$ because then $M_{\rm max} \lesssim 1$.  \par
The force multiplier depends also on other gas and radiation properties, for example gas metallicity (e.g., CAK) and column density, $N_{\rm H}$ \cite[e.g., ][]{Stevens91}. However, here we concentrate only on the effects due to $t$ and $\xi$ for a given SED and chemical abundance \cite[the same as the ones in][]{MKK16}.
\begin{figure*}
  \centering
  \includegraphics[width=\textwidth]{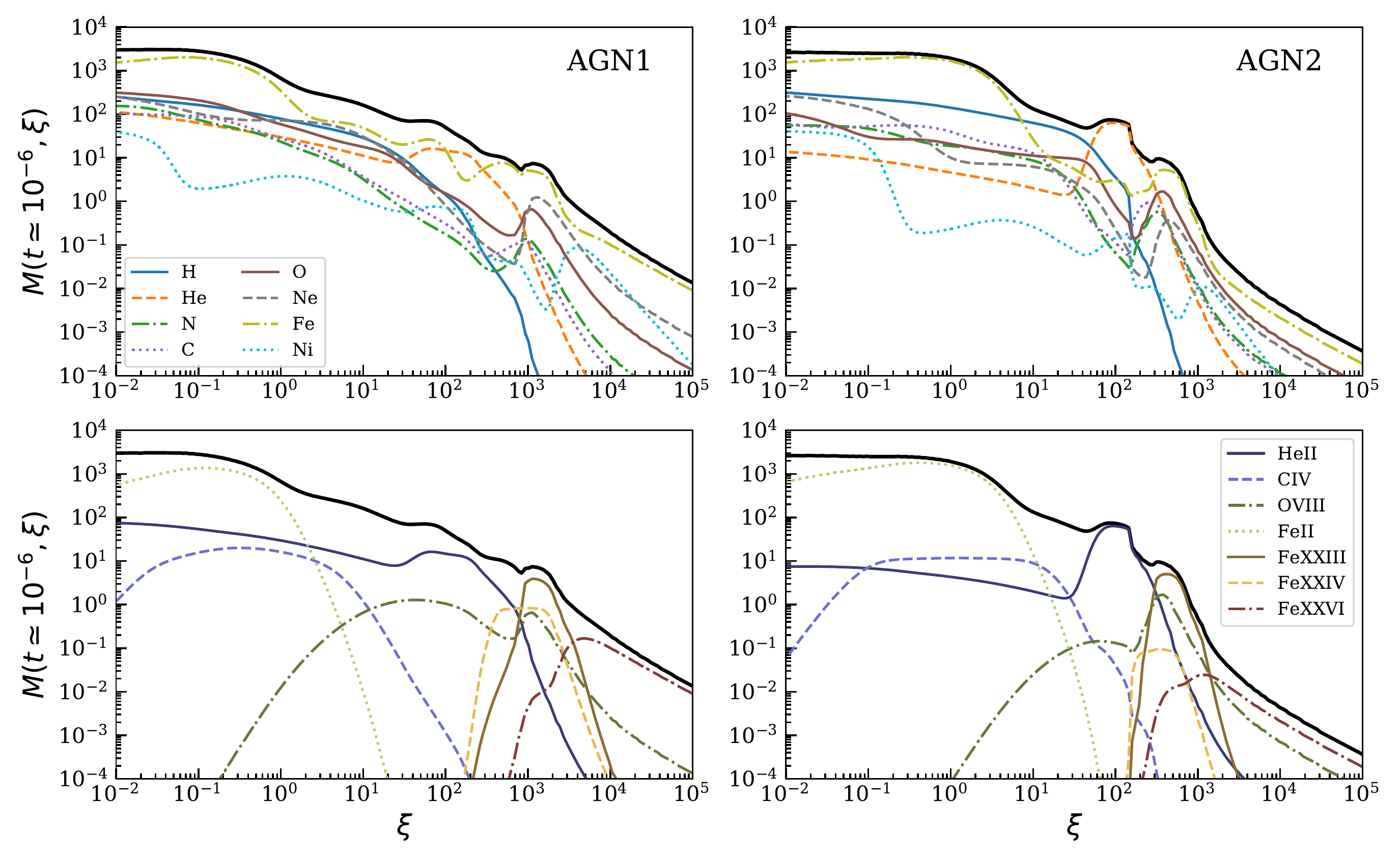}
  \caption{ Main contributors to the total force multipliers.  The left panels correspond to AGN1 and the right panels correspond to AGN2.  The solid black line in each panel represents the total force multiplier for $t\simeq10^{-6}$ (our proxy for $M_{\rm max}$).  The top panels shows the contribution to $M_{\rm max}$ for various elements of interest (see the legend in the bottom left corner of the top left panel). Similarly, the bottom panels show the contribution of various ions (see the legend in the bottom right corner of the bottom left panel).}
  \label{fig:mt_elem_ion}
\end{figure*}
\section{Results} \label{sec:results}
In Fig.~\ref{fig:summary-fig}, we show how the force multiplier changes as a function of $\xi$ and $t$. The left hand side panels show that $M(\xi,t)$ is a mostly monotonic function of $t$ that saturates at  very small $t$ for all $\xi$, while at large $t$, $M(\xi,t)$ is a power law of $t$ as  first found by CAK.  The right hand side panels show that $M(\xi,t)$ decreases slightly with increasing $\xi$ for small $\xi$ and a fixed $t$.  However, for $\xi \gtrsim 1$, $M(\xi,t)$ decreases significantly with increasing $\xi$ (the details depend on $t$).  The decrease is not monotonic; there is a resurgence in the line force for $\xi\approx 100-1000$ for both the AGN1 and AGN2 cases.  A similar ``bump'' feature on the plot of $M(\xi,t)$ vs. $\xi$ was shown in SK90 (see Fig.~2 there).  No such feature was shown in the work of \citet[][]{E05},  \citet[][]{Cetal09}, and \citet[][]{SC11}; their models indicate that there should be a dramatic decrease of $M(\xi,t)$ in that region. Our calculations are nevertheless in agreement with the finding made by these recent studies showing that $M(t,\xi)$ can be larger than 1000 for low values of $\xi$.\par
\begin{table*}
\centering
\begin{tabular}{ l l l !{\vrule width 2pt} l l l }
\hline
 & AGN1 & & & AGN2 & \\ \hline
 log($\xi$) $\simeq$ 0 & $M_{total}(t,\xi)=522$  & & 
 log($\xi$) $\simeq$ 0 & $M_{total}(t,\xi)=1635$ & \\ \hline 
  Ion & Wavelength (\angstrom) & $M_L(t,\xi)$ &
  Ion & Wavelength (\angstrom) & $M_L(t,\xi)$  \\  \hline		
  \ion{H}{1}  & 1216 & 5.051 & \ion{H}{1} & 1216 & 5.992 \\
  \ion{H}{1}  & 1026 & 4.768 & \ion{H}{1} & 1026 & 5.769 \\
  \ion{C}{4}  & 1550 & 1.550 & \ion{C}{4} & 1550 & 1.840 \\
  \ion{N}{5}  & 1239 & 1.355 & \ion{C}{3} & 977  & 1.562 \\
  \ion{O}{6}  & 1037 & 1.224 & \ion{N}{5}  & 1239 & 1.520 \\ 
  \hline  \hline
 log($\xi$) $\simeq$ 1 & $M_{total}(t,\xi)=123$ & &
 log($\xi$) $\simeq$ 1 & $M_{total}(t,\xi)=117$ & \\ \hline
  \ion{H}{1}  & 1216 & 4.771  & \ion{H}{1}  & 1216 & 7.522 \\
  \ion{Ne}{6} & 401  & 1.02   & \ion{H}{1}  & 1026 & 4.023 \\
  \ion{O}{6}  & 150  & 0.945  & \ion{C}{4} & 1548 & 2.227 \\
  \ion{O}{7}  & 21   & 0.724  & \ion{N}{5}  & 1239 & 1.907 \\
  \ion{C}{5}  & 40   & 0.6812 & \ion{O}{6}  & 1038 & 1.849 \\ 
  \hline  \hline
 log($\xi$) $\simeq$ 2 & $M_{total}(t,\xi)=24$ & &
 log($\xi$) $\simeq$ 2 & $M_{total}(t,\xi)=9.2$ & \\ \hline 
  \ion{O}{8}   & 19 & 0.604  & \ion{H}{1}   & 6563 & 0.409 \\
  \ion{Si}{10} & 303 & 0.399 & \ion{O}{6}   & 1032 & 0.171 \\
  \ion{Fe}{11} & 180 & 0.340 & \ion{Fe}{11} & 180  & 0.096 \\
  \ion{Fe}{12} & 202 & 0.330 & \ion{Fe}{10} & 174  & 0.089 \\
  \ion{Fe}{14} & 284 & 0.290 & \ion{Fe}{12} & 187  & 0.089 \\ 
  \hline  \hline
 log($\xi$) $\simeq$ 3 & $M_{total}(t,\xi)=20$ & &
 log($\xi$) $\simeq$ 2.6 & $M_{total}(t,\xi)=22$ & \\ \hline 
  \ion{Fe}{23}  & 777  & 7.328  & \ion{Fe}{23} & 1157 & 6.200 \\
  \ion{Fe}{22}  & 1157 & 3.546  & \ion{Fe}{23} & 1345 & 4.725 \\
  \ion{Fe}{22}  & 1344 & 2.640  & \ion{Fe}{23} & 3739 & 3.909 \\
  \ion{Fe}{22}  & 3738 & 2.172  & \ion{Fe}{23} & 9587 & 2.193 \\
  \ion{Fe}{22}  & 9586 & 1.561  & \ion{O}{8}   & 1303 & 0.416 \\ 
  \hline  \hline
\end{tabular}
\caption{Table showing the top five contributing lines to the force multiplier for AGN1 (left column) and AGN2 (right column) for fixed $t = 10^{-6}$ and selected values of $\xi$.  Note that the value of $\xi$ in the bottom section of the table differs in the each column; it is intended to correspond to the bump in the $M_{\rm max}$ distribution at large $\xi$ (see the black lines in fig.~\ref{fig:mt_elem_ion}).}
\label{table:mt-table}
\end{table*}
\begin{figure*}
  \centering
  \includegraphics[width=\textwidth]{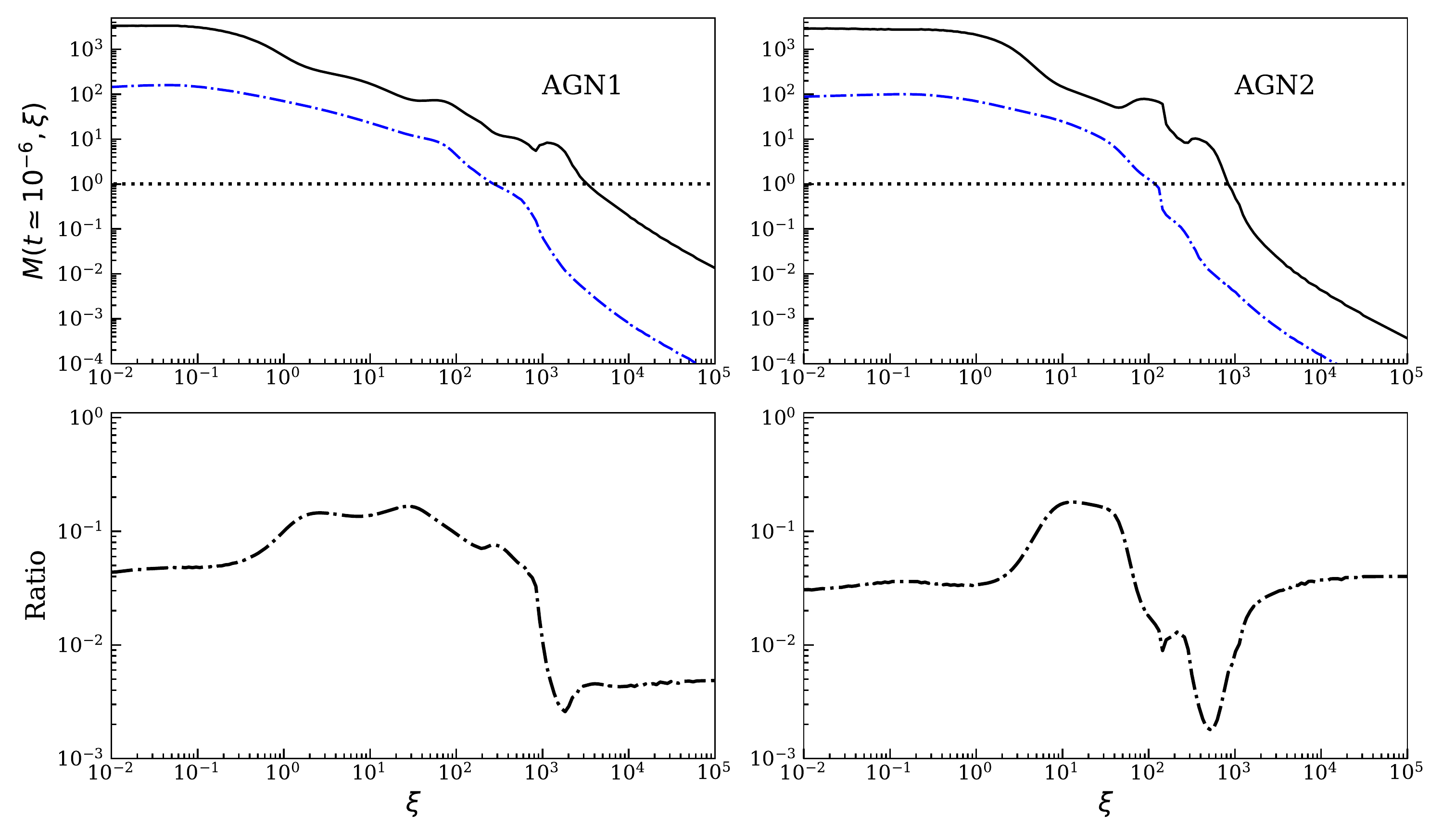}
  \caption{Comparison of LTE and non-LTE force multiplier calculations for AGN1 (left) and AGN2 (right), assuming $t=10^{-6}$.  \textit{Top Panels}:  The solid black line is $M(\xi,t)$ from Fig.~\ref{fig:mt_eta} that assumes a Boltzmann distribution for the level populations.  
  The blue dash--dotted line is $M(\xi,t)$ found without making the Boltzmann distribution assumption but with the level populations determined by \XSTAR~and including only lines contained in \XSTAR.
  \textit{Bottom Panels}: The black dash-dotted line is the ratio of the blue dash-dotted and black solid lines, illustrating how the combination of a smaller line list and a non-LTE treatment of the level populations reduces $M(\xi,t)$.}
  \label{fig:nlte-compare}
\end{figure*}
The force multiplier is a sum of opacities due to various lines. To better understand its properties,  in  Fig.~\ref{fig:mt_elem_ion} we show  the contributions to the force multiplier from various ions as a function of $\xi$ for $t=10^{-6}$.  We note that $M(t=10^{-6}, \xi)$ is our proxy for $M_{\rm max}$, so this figure illustrates how $M_{\rm max}$ changes with $\xi$.  
Similar to SK90, we find that Fe and He are the dominant contributors to the force multiplier.  Table~\ref{table:mt-table} lists the top five contributors to the force multiplier for four values of $\xi$.  The top contributors for low values of $\xi$ contribute $<5\%$ to the total force multiplier, agreeing with previous findings (e.g. \citealt{Abbott82}, \citealt{Gayley95}, \& \citealt{Puls00}).  Comparing to Fig.~\ref{fig:mt_elem_ion}, we notice that although He contributes a substantial amount to $M(\xi,t)$ at $\xi\simeq 100$, it is absent from our table.  Thus, this table illustrates an important result: the force multiplier is due to the contributions of many weak lines rather than a few very strong lines.
\par 
Returning to Fig. \ref{fig:mt_eta}, we demonstrate how $M(\xi,t)$ depends on the energy of the photons. We present the UV and X-ray contributions to the force multiplier (top panels). The UV band contributes the majority of the line force with the X-ray energy band only being important for a narrow window in AGN1 (i.e., for $\xi$ between 200 and 1000).  We find a similar trend for the  maximum opacity of a single line (bottom panels).  We note that even though opaque IR lines are present in our line list, we omit them from these two panels, because there are  comparatively few IR photons (see Fig.~\ref{fig:seds}) compared to UV and X-ray. The line opacity is weighted by the continuum flux, hence IR lines contribute very little to the total force multiplier.\par
We finish this section with a couple of observations based on a comparison of the results from the force multiplier calculations and the heating and cooling calculations that we carried out using the same code and input parameters. First, the significant increases of the equilibrium temperature with $\xi$ (see the solid black lines in the top panels of Fig. \ref{fig:mt_eta}) occurs for the ionization parameter range where $M_{\rm max}$ (solid blue lines) strongly decreases with increasing $\xi$. This is expected because spectral lines are major contributors to the gas cooling at small and intermediate $\xi$. Therefore, the decrease in the overall number of line transitions with increasing $\xi$ manifests itself in a decreases in $M(\xi,t)$ as well as an increase in the gas equilibrium temperature (i.e., an increase in the slope of the $\log~T$ vs $\log~\xi$ relation).  
\par
Second, the shaded regions in Fig. \ref{fig:mt_eta} mark the $\xi$ values for which the gas is thermally unstable (where the slope of the equilibrium curves exceed one). These regions closely coincide with those corresponding to the resurgence in the line force. As we will discuss in the next section, this overlap may strongly impact the significance of line driving for high ionization parameters where $M(\xi,t)$ is nominally  still larger than one.\par
\section{Discussion} \label{sec:discussion}
The basic requirement for line driving to win over gravity is $M_{\rm max} L' > 1$.  The non-monotonic behavior of $M_{\rm max}$ vs. $\xi$ implies that line-driving can be dynamically important over a relatively wide range of $\xi$, wider than explored in previous line-driven wind simulations, for example, those by \cite{PSK00}, \cite{PK04}, \cite{P07}, \cite{KP08}, \cite{KP09a}, \cite{KP09b} and \cite{Dyda18}. The line force might have been underestimated in those calculations.  We say this with caution, since the above requirement is just a necessary and not a sufficient condition for producing an appreciable line-driven wind.\par
Our finding that the gas can be thermally unstable over the same range of $\xi$ where there is a bump in the $M_{\rm max}$ distribution implies that this potential increase of the line force might not be physically realized.  This can be either because the flow avoids the thermally unstable region altogether (as in the 1D models of D17), or because cloud formation is triggered, thereby changing the local ionization state of the gas.  In either case, only a relatively small amount of gas (e.g., as measured by the column density) may be subject to the strong force at the location of the bump.  The effects of thermal instability in dynamical flows (\citealt{BS89}; \citealt{MP13}) must be further understood before a definitive conclusion can be drawn.  If cloud formation naturally occurs, the clouds will subsequently be accelerated (see Proga \& Waters 2015) and the resulting multiphase flow may have right properties (e.g., velocities and ionization structure) to account for the narrow line regions in AGN.  
\par
The bump at $\xi \approx 100-1000$ may also have interesting implications for the dynamics of line driven winds.  That is, a bump should be accompanied by different stages of acceleration, with distinct lines accompanying each stage, for Eddington fractions as low as 0.01. Note, however, that if the overall decrease in the number of lines, including coolants, leads to significant heating (i.e., runaway heating and thermal instability), the flow might be thermally driven rather than line-driven.  {Also not addressed are the effects of continuum opacity, which can contribute significantly to the total radiation force \citep[e.g.,][]{Stevens91,E05,SC11}.} 
\par
\subsection{LTE vs. non-LTE calculations}
One of the main challenges in estimating the force multiplier is the ability to properly compute the level populations, and although it is common to assume a Boltzmann distribution for the level populations (see \S\ref{sec:fmult-calc}), this assumption may not be appropriate for AGN, unjustifiably adding the contribution of thousands of additional lines.  One needs to have complete atomic data in order to solve the statistical equilibrium equations instead of assuming a Boltzmann distribution. To explore non-LTE effects, we used only the lines available in \XSTAR~because not all necessary atomic data is available for many lines included in Kurucz's list. \XSTAR~contains information about the vast majority of UV and X-ray lines included in our combined line list, and those are the lines present when the gas is highly ionized.
\par
The non-LTE calculation shown in Fig.~\ref{fig:nlte-compare} (blue dash-dotted line) is a demonstration of how a more complete treatment of the level populations reduces the force multiplier.  Our result that $M(\xi,t)$ can exceed unity for $\log(\xi)$ as large as 3 followed from the common practice of assuming a Boltzmann distribution for the level populations (see \S{\ref{sec:fmult-calc}}).
When we loosen this assumption by using the level populations determined by \XSTAR~ to compute $M(\xi,t)$ in a manner fully self-consistent with the heating and cooling rates, the population of the excited states is decreased.   The reason is simply that it is common for the temperature of photoionized plasmas to be an order of magnitude or more smaller than the ionization potential of H and He--like ions, so many excited state transitions that are collisionally populated in the LTE case can only exist if they become radiatively populated in the non-LTE case and radiative excitations can be highly sensitive to the incident SED. Consequently, the general expectation is a significant reduction of $M(\xi,t)$.
\par
Omitting lines that are exclusively present in Kurucz's list can also have a significant effect on the force multiplier, but only for low to moderate $\xi$.  We verified that for AGN1 with $\xi\gtrsim100$ and AGN2 with $\xi\gtrsim200$, there is no significant change in $M(\xi,t)$ when comparing Boltzmann distribution calculations done with and without the combined line list.  This means that non-LTE effects are most important for $\log(\xi)\gtrsim 2$, while for $\log(\xi)\lesssim 2$ we have found that non-LTE effects and the reduced line list contribute about equally to the $\sim 1$ order of magnitude reduction in $M(\xi,t)$. It is therefore likely that if we were able to perform a full non-LTE calculation using the combined line list, we would find that $M(\xi,t)$ is nearly as large as the black line in Fig.~\ref{fig:nlte-compare}, but only for $\log(\xi)\lesssim 2$.
\par
For $\log(\xi) \gtrsim 2$, the reduction in the line force is almost entirely due to non-LTE effects, and Fig.~\ref{fig:nlte-compare} shows $M(\xi,t)$ is smaller by more than two orders of magnitude, a finding at odds with our result that $M(\xi,t)$ can remain larger than unity out to $\log(\xi)\approx 3$.  This in itself is an important result, as the LTE approximation is a standard one when computing force multipliers (e.g., CAK; \citealt{Abbott82}; SK90; \citealt{Gayley95}; \citealt{Puls00}).  Note that it is dependent on the value of density we have assumed.  If densities are much greater than those considered in this work, the resulting $M(\xi,t)$ will be larger due to the greater populations of the excited levels due to electron collisions.
\par 
In consideration of our earlier discussion highlighting the fact that the gas is thermally unstable in the range $2 \lesssim \log(\xi) \lesssim 3$, the non-adiabatic hydrodynamics may dictate that force multipliers in this range of $\xi$ are seldom even sampled.  Understanding this highly ionized regime is likely impossible from the standpoint of photoionization physics alone; the gas dynamics must also be taken into account.
\par
\section{Conclusions} \label{sec:conclusion}
In the present paper, we have investigated the line force that results from exposing gas to AGN radiation fields.  In our calculations, we have used the most complete and up-to-date line list. We confirm the well-known result that the line force is a strong function of the photoniozation parameter for high values of the parameter and explore how this sensitivity depends on the AGN SED. We computed the force for several SEDs and here presented results for the SEDs that are representative of an obscured and unobscured AGN.  \par
These detailed force multiplier calculations have been made publicly available.  They assume a Boltzmann distribution for the energy levels, that the Sobolev approximation is valid, and that the optical depth between the gas incident SED is low ($N \leq 10^{17}~{\rm cm}^{-2}$) while the densities are low enough ($n_H = 10^{4}~{\rm cm}^{-3}$) such that collisional de-excitation processes are negligible compared to the radiation field in determining the ionization balance. The following is a summary of our results:
\begin{itemize}
    \item For a fixed value of the optical depth parameter, $t$, the force multiplier is not a monotonic function of the ionization parameter.  While $M(\xi,t)$ first decreases with $\xi$ for $\xi> 1$ as shown by others (e.g., SK90), this decrease is not as strong for both our obscured and unobscured  AGN SEDs, and in fact we found that $M(\xi,t)$ can increase again at $\xi \approx 30$ and $\xi \approx 900$ for AGN1 and $\xi \approx 50$ and $\xi \approx 300$ for AGN2.  The main consequence of this behavior is that the multiplier can stay larger than 1 for $\xi$ approaching $10^3$.  However, our non-LTE calculations in \S{4.1} revealed that non-LTE effects can reduce $M(\xi,t)$ by more than 2 orders of magnitude for $\log(\xi)\gtrsim 2$.  We pointed out that the range $10^2 \lesssim \xi \lesssim 10^3$ is also where gas is thermally unstable by the isobaric criterion, so fully understanding this highly ionized regime likely requires coupling photoionization physics with hydrodynamical calculations.
    \item Both the AGN1 and AGN2 SEDs lead to large force multipliers, but their $M(\xi,t)$ distributions differ significantly.  Therefore, in scenarios requiring the use of both obscured and unobscured SEDs in the same setting, for example when a significant amount of material crosses the line of sight between the source and observer, the gas dynamics will be affected.  If the incident SED suddenly transitions from AGN1 to AGN2, then this may lead to a decrease of $\xi$ by a factor of 3.4 (the ratio of the X-ray flux between the two SEDs assuming the same luminosity), and therefore both the heating and cooling rates and radiation force can be dramatically altered.
\end{itemize}
It is premature to state if our new results for the line force have significant consequences in explaining UV or X-ray absorbing outflows such as BALs, warm absorbers or ultra-fast outflows.  Therefore, in our follow-up paper, we will present results from our radiation-hydrodynamical simulations of outflows experiencing radiative heating and line-force from the same radiation flux. \par
\acknowledgments
We thank Sergei Dyda for useful discussions and our referee for encouraging exploration of non-LTE effects.  Support for Program number HST-AR-14579.001-A was provided by NASA through a grant from the Space Telescope Science Institute, which is operated by the Association of Universities for Research in Astronomy, Incorporated, under NASA contract NAS5-26555. This work also was supported by NASA under ATP grant NNX14AK44.\par \newpage

\end{document}